\documentclass[11pt,paper]{article}
\usepackage{jheppub}

\usepackage{float,slashed}
\usepackage[usenames,dvipsnames,table]{xcolor}
\usepackage{graphicx,amsmath,amssymb,multirow,array,bm,mathrsfs}
\usepackage{epsf,amsfonts,slashed}
\usepackage[numbers,sort&compress]{natbib}
\usepackage[export]{adjustbox}
\usepackage{tikz}
\usetikzlibrary{decorations.pathmorphing}
\usetikzlibrary{decorations.markings}
\usepackage{scalerel,stackengine} %Used to define a long hat
\usepackage{soul}
\usepackage[utf8]{inputenc}
\usepackage{hyperref}
\usepackage{tikz-feynman}

\newcommand{\AY}[1]{{\textcolor{Red}{ [AY:#1] }}}
\newcommand{\YO}[1]{{\textcolor{Cyan}{ [YO:#1] }}}

\title{
Holographic Turbulence From a Random Gravitational Potential 
}

\author{Yaron Oz,${}^{a}$,\,Sebastian Waeber${}^b$,\,Amos Yarom${}^b$}

\affiliation{$^a$
School of Physics and Astronomy, Tel-Aviv University, Tel-Aviv 69978, Israel
 }
 \emailAdd{yaronoz@tauex.tau.ac.il}
\affiliation{$^b$Department of Physics, Technion, Haifa 32000, Israel}

\emailAdd{wsebastian@campus.technion.ac.il}
\emailAdd{ayarom@physics.technion.ac.il}

\abstract{
We study the turbulent dynamics of a relativistic $(2+1)$-dimensional fluid placed in a stochastic gravitational potential. We demonstrate that the dynamics of the fluid can be obtained using a dual holographic description realized by an asymptotically Anti-de Sitter black brane driven by a random boundary metric. Using the holographic duality we study the inverse cascade
energy power spectrum of the fluid and show that it is compatible with that of a compressible fluid flow.
We calculate the local energy dissipation and the local fluid velocity distribution which provide other
measures of the holographic fluid turbulence.
}
\begin{document}

\maketitle

\section{Introduction}
Turbulence is a ubiquitous phenomenon associated with fluid motion. It is observed in an extremely wide variety of natural occurances, from the flow of blood through arteries, through the motion of swimmers, to atmospheric circulation and then, possibly, to stellar evolution. Moreover, due its robust nature, the study of turbulence impacts almost the entire spectrum of scientific research, from mathematics, to physics, to engineering. Nevertheless, despite its importance and impact, it is still not fully understood.

%In this work we further develop a recent proposal purporting to obtain a new perspective on turbulence by relating it to black hole physics. Such a relation between black holes and fluid dynamics has a somewhat lengthy history. The membrane paradigm relating the dynamics of event horizons to fluid flows was first discussed in \cite{1988SciAm.258d..69P}. This proposal has taken a somewhat circuitous path but has been made sharp within the context of the AdS/CFT correspondence \cite{Maldacena:1997re,Gubser:1998bc,Witten:1998qj}. In particular, thermal states of field theories with holographic duals can be described by AdS black hole geometries \cite{Witten:1998zw}, and therefore, fluid behavior of these theories can be captured by the dynamics of black holes \cite{Bhattacharyya:2007vjd}. Thus, it stands to reason that the turbulent behavior of fluids can be encoded in the geometry of black holes, i.e. in the horizon  \cite{Eling:2009pb,eling2010relativistic,eling2011gravity}.

This study endeavors to delve deeper into a recent proposal that seeks a novel perspective on turbulence by linking it to black hole physics. The interplay between black holes and fluid dynamics traces back to the membrane paradigm, initially expounded in \cite{1988SciAm.258d..69P}. Since its inception the membrane paradigm has been developed in various interesting directions but its sharpest manifestation is within the context of the AdS/CFT correspondence \cite{Maldacena:1997re,Gubser:1998bc,Witten:1998qj}. Notably, the thermal states of field theories with holographic duals are described by AdS black hole geometries \cite{Witten:1998zw},  implying that the fluid behavior of these theories can be encapsulated by the dynamics of black holes \cite{Bhattacharyya:2007vjd}. Thus, it stands to reason that the turbulent behavior of fluids can be encoded in the geometry of black holes \cite{Eling:2009pb,eling2010relativistic,eling2011gravity}.

Indeed, following the pioneering work of \cite{Adams:2013vsa} it was realized that the turbulent flow of fluids which posses holographic duals can be encoded in the geometric properties of dynamical black holes. Relations between scaling exponents of the velocity field and postulated fractal properties of the event horizon have been studied in \cite{Adams:2013vsa,Rozali:2017bll,Westernacher-Schneider:2017xie,Chen:2018nbh} and geometric aspects of vorticity were considered in \cite{Eling:2013sna,Green:2013zba,Yang:2014tla,Westernacher-Schneider:2015gfa,Marjieh:2021gln}. The focus of these latter works has been predominantly on decaying turbulence, where a random driving force is absent. As far as we know there are only a handful of studies where steady state turbulence was examined in a holographic context \cite{Ashok:2013jda,Balasubramanian:2013yqa,Westernacher-Schneider:2017xie,Andrade:2019rpn,Waeber:2021xba}. 

The current work aims at further exploring the construction of \cite{Waeber:2021xba} where turbulent flow was generated by placing a fluid in a stochastic background metric. 
Indeed, the authors of \cite{Waeber:2021xba} observed a Kolmogorov/Kraichnan scaling of the nonrelativistic energy power spectrum once the space-space components of the $(2+1)$-dimensional metric was set to fluctuate. Such a scaling behavior of the energy power spectrum is expected, on general grounds \cite{Kolmogorov1941,1967PhFl...10.1417K}, for nonrelativistic incompressible flow. Little is known regarding scaling behavior of %compressible turbulent flows in the non relativistic \cite{falkovich2010new,Wagner_2012,banerjee2014kolmogorov},
%and 
turbulent flows in the relativistic regime \cite{Fouxon:2009rd,Westernacher-Schneider:2015gfa,Eyink:2017zfz}. 

Given that the magnitude of the velocity field in \cite{Waeber:2021xba} is much smaller than the speed of light, $c$, and that a Kolmogorov/Kraichnan power spectrum is manifest, one might be lead to the conclusion that the flow described in \cite{Waeber:2021xba} matches that of an incompressible fluid. Such a conclusion might also be supported by general arguments given in  \cite{fouxon2008conformal,Bhattacharyya:2008kq,Eling:2009pb,Cai:2012vr} (see also \cite{Jensen:2014wha}) which discuss an appropriate scheme for obtaining incompressible fluid flow as an appropriate limit of a relativistic (and compressible) one. Neverhteless, %, as we discuss in this note, 
the nonrelativistic fluid flow observed by \cite{Waeber:2021xba} and its generalization is, in fact, compressible.

To set the stage for our computation, we review, in section \ref{S:NRlimit} the reduction of the relativistic fluid equations of motion to their nonrelativistic incompressible counterpart and also point out how a compressible component of the flow may, eventually, dominate the dynamics. We will also define the random gravitational potential and its required nonrelativistic scaling.
In section \ref{E:Setup} we present a holographic dual gravitational background
of a forced turbulent relativistic fluid flow, where the time-time component of the background metric being  a stochastic variable. The nonrelativistic limit is a fluid flow subject to a random 
force described holographically by a fluctuating gravitational potential. In section \ref{E:Results}
we provide numerical results demonstrating the turbulent nature of the resulting flow. 
We end with a discussion of our results in section \ref{E:Discussion}.

\section{The Nonrelativistic Limit}
\label{S:NRlimit}

Consider a relativistic uncharged conformal fluid characterized by a velocity field $u^{\mu}$ and temperature field $T$, moving in a background metric $g_{\mu\nu}$ with $\mu=0,1,2$ spacetime indices. The equations of motion for such a fluid are given by 
\begin{equation}
\label{E:conservation}
	\nabla_{\mu}T^{\mu\nu}=0 \ ,
\end{equation}
where 
\begin{equation}
\label{E:constitutive}
	T^{\mu\nu} = e(T) u^{\mu}u^{\nu} + P(T) P^{\mu\nu} -  2 \eta(T) \sigma^{\mu\nu} - \zeta(T) P^{\mu\nu} \nabla_{\alpha}u^{\alpha} + \mathcal{O}(\nabla^2)\,,
\end{equation}
and we have defined the projection
\begin{equation}
	P^{\mu\nu} = u^{\mu}u^{\nu} + g^{\mu\nu}\,,
\end{equation}
and the shear tensor,
\begin{equation}
	\sigma^{\mu\nu} =\frac{1}{2} P^{\mu\alpha}P^{\nu\beta} \left(\nabla_{\alpha}u_{\beta}+\nabla_{\beta}u_{\alpha}-g_{\alpha\beta}\nabla_{\lambda}u^{\lambda}\right) \,.
\end{equation}
The scalar functions $e$, $P$, $\eta$ and $\zeta$ are referred to as the energy density, the pressure, the shear viscosity and the bulk viscosity and they depend only on the temperature, $T$.

The explicit form of the equations of motion \ref{E:conservation} is given by
\begin{subequations}
\label{E:RNS}
\begin{equation}
\label{E:RNSscalar}
	\frac{1}{c_s^2} u^{\mu}\nabla_{\mu} \ln T  + \nabla_{\mu} u^{\mu} = \frac{2 \eta}{s T} \sigma_{\mu\nu}\sigma^{\mu\nu} + \frac{\zeta}{s T} \left(\nabla_{\mu}u^{\mu}\right)^2 + \mathcal{O}(\nabla^3) 
\end{equation}
and
\begin{multline}
\label{ERNSvector}
	u^{\mu}\nabla_{\mu}u^{\sigma} + P^{\sigma\mu} \nabla_{\mu} \ln T = P^{\sigma\mu}\left(\frac{2\eta}{s T}\nabla_{\nu}\sigma_{\mu}^{\nu} +  2\frac{\partial \eta}{\partial P}\sigma_{\mu\alpha}\nabla^{\alpha} \ln T\right) \\
	+ 
	\frac{\zeta}{s T} \left( \nabla_{\mu}u^{\mu} u^{\alpha}\nabla_{\alpha} u^{\sigma} + P^{\sigma\alpha}\nabla_{\alpha} \nabla_{\mu}u^{\mu} \right) + \frac{\partial \zeta}{\partial P} P^{\sigma\alpha} \nabla_{\mu}u^{\mu} \nabla_{\alpha} \ln T
	+
	\mathcal{O}(\nabla^3)\ ,
\end{multline}
\end{subequations}
where
\begin{equation}
	c_s^2 = \frac{\partial P}{\partial e}
\end{equation}
is the speed of sound squared and we have used
\begin{equation}
	e + P = s T
\end{equation}
with $s = \frac{\partial P}{\partial T}$ the entropy density.
The equations of motion \eqref{E:RNS} should be thought of in terms of a derivative expansion. The left hand side of \eqref{E:RNS} represents the leading order (ideal) terms and the right hand side of \eqref{E:RNS} represents the subleading (viscous) terms. Order $\mathcal{O}(\nabla^3)$ terms have been neglected.
%The equations of motion \eqref{E:RNS} have been written in terms of a derivative expansion have terms of a derivative expansion the left hand side of \eqref{E:RNS} are the leading terms while the right hand side of \eqref{E:RNS} is associated with dissipation. 

In what follows we will discuss the non relativistic limit of these equations in the presence of an external background metric. In section \ref{SS:NRL} we will consider the nonrelativistic limit of \eqref{E:RNS} and argue that if the driving has an irrotational component, the behavior of the fluid in the small velocity limit will be compressible. In section \ref{SS:Driving} we will focus more specifically on stochastic driving of the fluid via a boundary metric.

\subsection{Nonrelativistic Limit of Conformal Hydrodynamics}
\label{SS:NRL}

To gain a better understanding of the structure of \eqref{E:RNS} let us use $u^{\mu} = \frac{1}{\sqrt{1-\beta^2}} \left( 1, \vec{\beta} \right)$ (and assume a Minkowski space background) in which case \eqref{E:RNS} takes the form:
\begin{align}
\begin{split}
\label{E:REuler}
	\partial_t{\ln T} &= 
	-\frac{1}{1-c_s^2 \beta^2}\left((1-c_s^2)\beta \cdot \partial \ln T + c_s^2 \partial \cdot \beta \right) +\mathcal{O}(\nabla^2) \\
%	\partial_t{\beta}^i & = \frac{(1-\beta^2)\beta^i}{(1-c_s^2\beta^2)}\left((1-c_s^2)\beta\cdot \partial \ln T + c_s^2 \partial \cdot \beta\right) - \left(\beta \cdot \partial \beta^i + (1-\beta^2) \partial^i \ln T\right)+\mathcal{O}(\nabla^2) \\
	\partial_t{\beta}^i & = -(1-\beta^2)\beta^i \partial_t \ln T - \left(\beta \cdot \partial \beta^i + (1-\beta^2) \partial^i \ln T\right)+\mathcal{O}(\nabla^2) \ . 
\end{split}
\end{align}

To understand the incompressible limit of \eqref{E:REuler}, it is useful to consider the behaviour of \eqref{E:REuler} at long time and length scales. To this end, we define \cite{fouxon2008conformal}:
\begin{equation}
	\ln T(\vec{x},t) =\ln \mathcal{T}_0 + \epsilon^2 \mathcal{T}(\epsilon^n \vec{x},\epsilon^{n+1} t)
	\qquad
	\beta^i(\vec{x},t) = \epsilon v^i(\epsilon^n \vec{x},\epsilon^{n+1} t)
\end{equation}
with $\mathcal{T}_0$ a constant. Later we intend to take the $\epsilon \to 0$ limit while keeping $\tau=\epsilon^{n+1} t$ and $\chi = \epsilon^n x$ constant. The resulting equations take the form
\begin{align}
\begin{split}
\label{E:RNSepsilon}
	c_s \partial \cdot v 
	& = -\epsilon^2 \left( (1-\epsilon^2 c_s^2 v^2) \partial_{\tau}\mathcal{T} + (1-c_s^2)v \cdot \partial \mathcal{T} \right) +\mathcal{O}(\nabla^2) \\
%	\partial_t{\beta}^i & = \frac{(1-\beta^2)\beta^i}{(1-c_s^2\beta^2)}\left((1-c_s^2)\beta\cdot \partial \ln T + c_s^2 \partial \cdot \beta\right) - \left(\beta \cdot \partial \beta^i + (1-\beta^2) \partial^i \ln T\right)+\mathcal{O}(\nabla^2) \\
	\partial_\tau{v}^i + v \cdot \partial v^i + \nabla^i \mathcal{T} & = \epsilon^2 \left( - v^i (1-\epsilon^2 v^2) \epsilon^2 \partial_\tau \mathcal{T} - v^2  \partial^i \mathcal{T}\right) +\mathcal{O}(\nabla^2) \,.
\end{split}
\end{align}
Note that \eqref{E:RNSepsilon} does not depend on $n$. Once we consider the $\mathcal{O}(\nabla^2)$ viscous corrections to \eqref{E:REuler} we will set $n=1$. Without any further assumption, equations \eqref{E:RNSepsilon} hold the same information as \eqref{E:REuler}. 

If we take the $\epsilon \to 0$ limit of \eqref{E:RNSepsilon} while keeping $\tau$ fixed we obtain the unsourced Euler equations for a dissipationless fluid. In this limit, the dynamical equation for $\mathcal{T}$ (which represents the pressure of the fluid) becomes a constraint equation for the velocity $v^i$. By taking the $\tau$ derivative of the constraint equation and the divergence of the dynamical equation for $v^i$ one obtains a Poisson equation for $\mathcal{T}$. Thus, when $\epsilon=0$, the resulting initial value problem requires data regarding $v^i$ at some initial value of $\tau$.  Equations \eqref{E:RNSepsilon} suggest that if the velocity is sufficiently small at late times, then, independently of the initial value provided for the velocity field and pressure, the late time solution will converge to that of the Euler equation.

It is straightforward to add $\mathcal{O}(\nabla^2)$ viscous corrections to \eqref{E:RNSepsilon} and also place it on a manifold with a slightly perturbed background metric. Indeed, following \cite{Bhattacharyya:2008kq,Eling:2009pb} one finds
\begin{align}
\begin{split}
\label{E:INS}
	\partial \cdot v &= \mathcal{O}(\epsilon^2,\,\nabla^3) \\
	\partial_{\tau} v_i + v\cdot \nabla v^i + \nabla^i \mathcal{T} - \nu \nabla^2 v^i &= f^i + \mathcal{O}(\epsilon^2,\,\nabla^3)
\end{split}
\end{align}
with
\begin{equation}
	\nu = \frac{ 2}{T_0} \frac{ \eta(T_0)}{s(T_0)} \,,
	\qquad
	f^i = -\nabla^i \varphi - \partial_t A^i - F^{ij} v_j \ ,
\end{equation}
and $F = dA$ and we have used
\begin{equation}
\label{E:scalingmetric}
	g_{\mu\nu}dx^{\mu}dx^{\nu}  = \eta_{\mu\nu}dx^{\mu}dx^{\nu}  - 2 \epsilon^2 \varphi(\chi,\tau) dt^2 + 2 \epsilon A_i (\chi,\tau) dx^i dt + \epsilon^2 h_{ij}(\chi,\tau) dx^i dx^j \ ,
\end{equation}
and set $n=1$. %Note that we may always tune our background source such that it will satisfy the scaling relations in \eqref{E:scalingmetric} and one might then expect that in the small $\epsilon$ limit the incompressible Navier-Stokes equation \eqref{E:INS} with $\epsilon=0$ will be satisfied. 
We will now argue that, with this scaling, if the driving force, $f^i$, is conservative then the flow will never be incompressible at late times for any finite value of $\epsilon$.

We refer to the $\epsilon = 0$ limit of \eqref{E:INS} as the strictly incompressible fluid, viz., one that satisfies
\begin{equation}
\label{E:strict}
	\partial_{\tau}\vec{u} + \vec{u} \cdot \vec{\nabla} \vec{u} = - \vec{\nabla} p + \nu \nabla^2 \vec{u} +\vec{\nabla}\varphi \,,
	\qquad
	\vec{\nabla} \cdot \vec{u} = 0\,.
\end{equation}
(Here, we have used $\vec{u}$ instead of $\vec{v}$ as the velocity field, and $p$ instead of $\mathcal{T}$ for the pressure in order to make it clear that we are in the strictly incompressible limit.)
To better understand the structure of \eqref{E:strict} it is convenient to carry out a Helmholtz decomposition of the velocity field. That is, given a vector $\vec{w}$ we define its potential $\phi_w$ via
\begin{equation}
	\nabla^2 \phi_w = \vec{\nabla} \cdot \vec{w}\,.
\end{equation}
The potential $\phi_w$ is defined up to a harmonic function. On a torus $\phi_w$ is defined up to a constant so $\vec{\nabla} \phi_w$ is unique. We now define the compressible part of $\vec{w}$,
\begin{equation}
	\vec{\Pi}_c(\vec{w}) = \vec{\nabla} \phi_w \ ,
\end{equation}
which is uniquely defined on the torus. We define the incompressible part of $\vec{w}$ as
\begin{equation}	
	\vec{\Pi}_i(\vec{w}) = \vec{w} - \vec{\Pi}_c (\vec{w})\,.
\end{equation}
Acting with $\Pi_i$ and $\Pi_c$ on \eqref{E:strict} we find
\begin{align}
\begin{split}
\label{E:strictcomponents}
	\partial_{\tau} \vec{u}_c + \vec{\Pi}_c ( \vec{u} \cdot \vec{\nabla} \vec{u}) &= - \vec{\nabla}p + \nu \nabla^2 \vec{u}_c + \vec{\nabla}\varphi \\
	\partial_{\tau} \vec{u}_i + \Pi_i(\vec{u} \cdot \vec{\nabla} \vec{u}) &= \nu \nabla^2 \vec{u}_i  \\
	\vec{\nabla} \cdot \vec{u}_c & = 0 \ ,
\end{split}
\end{align}
where we have defined
\begin{equation}
\label{E:udecomposition}
	\vec{u}_i = \Pi_i(\vec{v})\,,
	\qquad
	\vec{u}_c = \Pi_c(\vec{v})\,.
\end{equation}

The last equality in \eqref{E:strictcomponents} implies that $\vec{u}_c = \vec{\nabla} \phi_{u_c} = 0$. Implementing this in \eqref{E:strictcomponents} and writing $p' = p - \Phi$ we find that \eqref{E:strictcomponents} reduces to
\begin{align}
\begin{split}
\label{E:strictcomponents2}
	\vec{\Pi}_c ( \vec{u} \cdot \vec{\nabla} \vec{u})\Big|_{\vec{u}_c=0} &= - \vec{\nabla}p'  \\
	\partial_{\tau} \vec{u}_i + \vec{\Pi}_i(\vec{u} \cdot \vec{\nabla} \vec{u})\Big|_{\vec{u}_c=0}  &= \nu \nabla^2 \vec{u}_i + \vec{f}_i \ .
\end{split}
\end{align}
Note that since $p'$ is determined by the first equation in \eqref{E:strictcomponents2}, given initial conditions for $\vec{u}_i$ on some initial time slice the evolution of the velocity field will be independent of the potential $\Phi$. Since equations \eqref{E:strictcomponents} are equivalent to the strictly incompressible Navier Stokes equation we will assume that at late times the velocity and temperature will reach a thermally equilibrated configuration. In other words, given some initial data $\vec{u}_i(\chi,0)=\vec{u}_0(\chi)$ (for which the total momentum of the fluid is zero) there will exist some late time $T$ after which $|\vec{\nabla}p|$, $|\dot{p}|$ and $|\vec{u}_i|$ are sufficiently small.

Now let us consider the dynamics associated with a driven compressible fluid. The associated equations of motion are an extended version of \eqref{E:RNSepsilon},
\begin{align}
\begin{split}
	c_s \vec{\nabla} \cdot \vec{v} & = -\epsilon^2 \left( \partial_{\tau} \mathcal{T} + \left( \substack{\hbox{terms which involve products} \\ \hbox{of more than one field} }\right) \right)+ \mathcal{O}(\nabla^3) \\
	\partial_{\tau}\vec{v} + \vec{v} \cdot \vec{\nabla} \vec{v} &= - \vec{\nabla} \mathcal{T} + \nu \nabla^2 \vec{v} +\vec{\nabla}\varphi + \epsilon^2 \left( \substack{\hbox{terms which involve products} \\ \hbox{of more than one field} }\right) + \mathcal{O}(\nabla^3)\,.
\end{split}
\end{align}
By ``terms which involve products of more than one field'' we mean expressions which are at least quadratic in $\vec{v}$, $\mathcal{T}$ and $\varphi$. We point out, once again, that $\mathcal{T}$ plays the role of pressure. Now, carrying out the same analysis that lead to \eqref{E:strictcomponents2} we find
\begin{align}
\begin{split}
\label{E:compressiblecomponents}
	\partial_{\tau} \vec{v}_c + \vec{\Pi}_c ( \vec{v} \cdot \vec{\nabla} \vec{v}) &= - \vec{\nabla}\mathcal{T}' + \nu \nabla^2 \vec{v}_c + \epsilon^2 \left(\ldots \right) \\
	\partial_{\tau} \vec{v}_i + \vec{\Pi}_i(\vec{v} \cdot \vec{\nabla} \vec{v}) &= \nu \nabla^2 \vec{v}_i +  \epsilon^2 \left(\ldots \right) \\
	\vec{\nabla} \cdot \vec{v}_c & = -\epsilon^2 \left( \partial_{\tau} \mathcal{T}' + \partial_{\tau} \varphi + \ldots \right)
\end{split}
\end{align}
with $\mathcal{T}' = \mathcal{T}-\varphi$. 

Let us assume, by contradiction, that the late time behavior of \eqref{E:compressiblecomponents} is given by the $\epsilon \to 0$ limit of \eqref{E:compressiblecomponents}, i.e., \eqref{E:strictcomponents}. That is, after some time, we have $\delta = |\vec{v}_c| \ll |\vec{v}_i|$. Since $|\vec{v}_i|$ can be made very small at late times, then also $\delta$ will be sufficiently small. Now, the last equation in \eqref{E:compressiblecomponents} reads
\begin{equation}
	\mathcal{O}(\delta) = \vec{\nabla} \cdot \vec{v} = -\epsilon^2 \partial_{\tau}\varphi + \mathcal{O}(\epsilon^2 \delta) \ ,
\end{equation}
which can not be balanced unless $\varphi$ is parametrically small or $\epsilon=0$. Thus, we conclude, that for finite values of $\epsilon$, the flow will never be incompressible at late times. (Somewhat more formally, the limits $\delta \to 0$ and $\epsilon \to 0$ don't commute.) 

To conclude, even though driving a strictly incompressible fluid with a conservative force will not modify the dynamics of the velocity, forcing any type of compressible flow will be felt by the velocity field. In this work we will use a random gravitational potential to generate compressible turbulent flow.

\subsection{A Random Gravitational Potential}
\label{SS:Driving}
The main goal of this work is to study the manifestation of turbulent flow due to a stochastic gravitational potential in black brane geometries. Indeed, steady state turbulent flow requires a random driving force, which would continuously inject energy into the system at a particular length (or momentum) scale which we will refer to as the driving scale. In a direct cascade this energy is dissipated at the much lower viscous length scale, while in an inverse cascade this energy is dissipated at larger scales by other means, such as large scale friction. 

There are many methods of generating a random gravitational potential. In what follows we will follow the method discussed in \cite{Waeber:2021xba} which has been proved useful in a holographic context. We choose $ds^2 = -\left( 1 + 2 \Phi\right)dt^2 + \sum_i (dx^i)^2 $ such that
\begin{equation}
\label{E:Qtoq}
	2 \Phi = q(t,\vec{x}) + 3\, \sqrt{ \overline{ q(t,\vec{x})^2 } } \ ,
\end{equation}
where $q(t,\vec{x})$ is a random variable with zero mean which we will specify shortly, and barred quantities denote an average. Thus $\sqrt{ \overline{ q(t,\vec{x})^2 } }$ denotes the variance of $q$. %The reason we have added three times the variance of $q$ to $\Phi$ is historical and will be updated in future work.%that we want statistical fluctuations of $\Phi$ to deviate only slightly from its mean. 

In order to have well defined time derivatives of $\Phi$, we define $q$ through the stochastic differential equation
\begin{equation}
\label{E:dotq}
	\frac{\partial}{\partial t}  q (t,\vec{x}) = - \frac{q(t,\vec{x})}{\tau} + \frac{\xi(t,\vec{x})}{\tau}\,,
\end{equation}
using Stratanovich evolution. Here $\tau >0$ is referred to as the autocorrelation time, and $\xi$ is given by white noise distribution in time with zero average and whose variance, parameterized by $D>0$, ranges, in Fourier space, over an annulus $A$ of inner radius $k_f-\Delta k$ and outer radius $k_f+\Delta k$:
\begin{equation}
\label{E:averagexi2}
	\overline{\xi(t,\vec{x})} = 0\,,
	\qquad
	\overline{\xi(t,\vec{x})\xi(t',\vec{x}')} = D \delta(t-t') \sum_{\substack{i \\ {\vec{k}_i \in A}}} \cos\left(\vec{k}_i \cdot \left(\vec{x}-\vec{x}'\right) \right)\,.
\end{equation}
Higher moments of $\xi$ are given by Isserlis' theorem (Wick's theorem), e.g., $\overline{\xi_1 \xi_2 \xi_3 \xi_4 } =  \overline{\xi_1 \xi_2} \, \overline{\xi_3 \xi_4} + \overline{\xi_1 \xi_3} \, \overline{\xi_2 \xi_4} + \overline{\xi_1 \xi_4} \, \overline{\xi_2 \xi_3}$.
Correlations of the form \eqref{E:averagexi2} can be obtained from the distribution 
\begin{equation}
\label{E:xidistribution}
	\xi(t,\vec{x}) = \sqrt{D}\sum_{\substack{i \\ {\vec{k}_i \in A}}} \left( P^{(1)}_{i}(t) \cos \left(  \vec{k}_i \cdot \vec{x} \right)+P^{(2)}_{i}(t)\sin \left( \vec{k}_i \cdot \vec{x} \right)\right) \ ,
\end{equation}
where the $P^{(a)}_{i}(t)$ satisfy $\overline{P^{(a)}(t)} = 0$ and $\overline{P ^{(a)}(t)P^{(b)}(t') } = \delta^{ab} \delta(t-t')$ and higher moments are given by Isserlis' theorem. (One way to generate the moments of such a distribution is to discretize time into steps $t_n = n \Delta t$,
then, for each $t_n$ and for each $a=1,2$ and $\vec{k}_i \in A$, draw $P^{(a)}_{i}(t)$ from a normal distribution with average zero and variance $1/\sqrt{\Delta t}$. Moments of $P^a(t)$ in the continuum can be defined by taking the continuum limit of moments of $P(t_n)$.)

Thus, we characterize the random force by 3 parameters, its strength, $D$, its autocorrelation time $\tau$, and the driving frequency, $k_f-\Delta k < k < k_f+\Delta k_f$. To better understand the role of these parameters, we note that the solution to \eqref{E:dotq} with initial conditions such that $q(0)=0$ is
\begin{equation}
	q(t,\vec{x}) = \frac{1}{\tau} \int_0^t e^{-\frac{t-t'}{\tau}} \xi(t',\vec{x}) dt'\,.
\end{equation}
Odd moments of $q$ obviously vanish and even moments can be read off from
\begin{equation}
\label{E:correlator}
	\overline{q(t_1,\vec{x}_1) q(t_2,\vec{x}_2)} = \frac{D}{2\tau}  \sum_{\substack{i \\ {\vec{k}_i \in A}}} \cos(\vec{k}_i \cdot \left(\vec{x}_1-\vec{x}_2)\right) \left(e^{\frac{|t_1-t_2|}{\tau}} - e^{-\frac{t_1+t_2}{\tau}}\right) \ ,
\end{equation}
and Isserlis' theorem. Likewise, moments of $\dot{q}(t)$ with $q(t_1)\ldots q(t_n)$ at unequal times can be derived from
\begin{equation}
	\overline{\dot{q}(t_1,\vec{x}_1) q(t_2,\vec{x}_2) \ldots q(t_n,\vec{x}_n)} = - \frac{1}{\tau} \overline{q(t_1,\vec{x}_1)q(t_2,\vec{x}_2)\ldots q(t_n,\vec{x}_n) }\,, \qquad t_1 > t_n \quad \forall n\geq 2\,.
\end{equation}
Also, taking spatial derivatives of $q$ commutes with averaging.
Thus, we associated the typical size of fluctuations associated with $q$ to be $\sqrt{\frac{D N_A}{2\tau}}$, where 
\begin{equation}
	N_A = \sum_{\substack{i \\ {\vec{k}_i \in A}}} 1\,
\end{equation}
is the ``volume'' of the annulus.
The time derivative of $q$ is of order $\frac{1}{\tau}$ and its spatial derivative is of order $k_f$ (assuming $\Delta k$ is sufficiently small).

In \eqref{E:scalingmetric} we assumed that
\begin{equation}
	\Phi(t,\vec{x}) = \epsilon^2 \varphi(\epsilon^2 t ,\,\epsilon x)
\end{equation}
such that $\epsilon$ is small and $\varphi$ is of order $\mathcal{O}(\epsilon^0)$. Thus, in order that
\begin{equation}
	\Phi \sim \mathcal{O}(\epsilon^2)\,,
	\qquad
	\partial_t \Phi \sim \mathcal{O}(\epsilon^4)\,,
	\qquad
	\partial_i \Phi \sim \mathcal{O}(\epsilon^3)\,,
\end{equation}
we should set
\begin{equation}
	\sqrt{\frac{D}{\tau} } \sim \mathcal{O}(\epsilon^2)\,,
	\qquad
	\frac{1}{\tau} \sqrt{\frac{D}{\tau}} \sim \mathcal{O}\left(\epsilon^4\right)\,,
	\qquad
	k_f  \sqrt{\frac{D}{\tau}} \sim \mathcal{O}(\epsilon^3)\,.
\end{equation}
The first approximation follows from \eqref{E:correlator}, the second from dimensionless analysis,\footnote{The quantity $\overline{\dot{q}\dot{q}}$ is not well defined. In order to make it well defined we would need to define $q$ through a cascade of Ornsetein-Uhlenbeck processes \cite{Waeber:2021xba}.} and the last equality from \eqref{E:averagexi2}. It follows that
\begin{equation}
	\frac{1}{\tau} \sim \epsilon^2\,,
	D \sim \epsilon^2\,,
	k_f \sim \epsilon\,.
\end{equation}
These parameters ensure that the source has the appropriate scaling so as to lead to a nonrelativistic limit of the form described in section \ref{S:NRlimit}. %As we will see shortly, we will see that this non relativistic flow is compressible (as predicted in section \ref{S:NRlimit}). 

\section{The Holographic Dual Gravitational Background}
\label{E:Setup}

%\subsection{Setting Up the Holographic Problem}
In the previous section we have argued that if a compressible fluid is acted on by a time dependent conservative force then the resulting flow will not be compressible. Consequentially, placing a fluid in a randomly fluctuating gravitational potential can not lead to incompressible turbulent flow and one may speculate on the emergent behavior of the fluid.

A priori, a natural framework to study compressible fluids in a fluctuating gravitational potential would be that of relativistic fluids whose equations of motion were given in \eqref{E:RNS}. Unfortunately, viscous relativistic fluid dynamics is, in a sense, causal and unstable \cite{PhysRevD.35.3723,ISRAEL1976310} and while there have been a variety of attempts at resolving these issues, e.g., \cite{ISRAEL1976213,HISCOCK1983466,Baier:2007ix,Denicol:2012cn,Bemfica:2017wps,Kovtun:2019hdm}, none are entirely satisfactory \cite{hydroworkshop}. That said, there exists a family of (conformal) fluids for which the evolution equations are stable and causal. These are fluids which possess a dual description in terms of black hole dynamics \cite{Bhattacharyya:2007vjd}. 

Following \cite{Maldacena:1997re,Gubser:1998bc,Witten:1998qj}, solutions to the equations of motion following from the variation of
\begin{equation}
\label{E:action}
	S =- \frac{1}{16\pi G} \int \sqrt{-g} \left(R -\frac{(d-2)(d-1)}{\ell^2}\right)  d^dx + \left(\substack{\hbox{ boundary} \\ \hbox{terms} }\right)
\end{equation}
correspond to states in a dual conformal field theory. More precisely, if we write the solution to the equations of motion following from \eqref{E:action} in a Fefferman Graham coordinate system,
\begin{equation}
\label{E:FGsol}
	ds^2 = \frac{\ell^2}{r^2}\left(dr^2 + g_{\mu\nu} dx^{\mu}dx^{\nu}\right) \ ,
\end{equation}
(where the $\mu$ indices do not cover the ``radial'' coordinate $r$), then the expectation value of the stress tensor in the state, $\rho$, dual to the solution is given by
\begin{equation}
\label{E:Tmap}
	\hbox{Tr} \left( \rho T^{\mu\nu} \right)_{g{(0)} }= \frac{(d-1)\ell^{d-2}}{16 \pi G_N} g_{(d)\,\mu\nu} + G_{\mu\nu}[g_{(0)}] \ ,
\end{equation}
where $g_{(n)}$ corresponds to an appropriate term in a series expansion of $g_{\mu\nu}$ near the asymptotic boundary located at $r=0$,
\begin{equation}
	g_{\mu\nu} = \sum_{n=0}r^{n}  g_{(n)\mu\nu}  + \mathcal{O}(r^{d-1} \ln r) \ .
\end{equation}
The subscript $g_{(0)}$ on the left hand side of \eqref{E:Tmap} indicates that the state is evaluated in a coordinate system associated with the metric $g_{(0)\mu\nu}$ on which the field theory is defined. The expression for $G_{\mu\nu}$ on the right hand side of \eqref{E:Tmap} is scheme dependent \cite{Buchel:2012gw}. According to the scheme of \cite{deHaro:2000vlm} it is, for $d=5$, given by
\begin{equation}
	G_{\mu\nu}[g_{(0)}] = -\frac{1}{8} g_{(0 \mu\nu} \left(\left(\hbox{Tr} g_{(2)}\right)^2 - Tr g_{(2)}^2 \right) - \frac{1}{2} g_{(2)\mu\alpha}g_{(2)}{}^{\alpha}{}_{\nu} + \frac{1}{4} g_{(2)\mu\nu} \hbox{Tr}g_{(2)}\,,
\end{equation}
where
\begin{equation}
	g_{(2)\mu\nu} = \frac{1}{d-3} \left(R_{(0)\mu\nu} - \frac{1}{2(d-2)} R_{(0)g_{(0)\mu\nu}}\right)
\end{equation}
with $R_{(0)\mu\nu}$ and $R_{(0)}$ the Ricci tensor and Ricci scalar associated with $g_{(0)\mu\nu}$ respectively. For $d=4$ the functional $G_{\mu\nu}$ vanishes \begin{equation}
    G_{\mu\nu}[g_{(0)}] = 0.
\end{equation}

For example, the black hole geometry, in Eddington-Finkelstein coordinates
\begin{equation}
\label{E:BH}
	\ell^{-2} ds^2 = \frac{1}{\rho^2} \delta_{ij} dx^i dx^j - \frac{1}{\rho^2} \left(1-\left(\frac{\rho}{\rho_0}\right)^{d-1}\right) dt^2 - \frac{2 dt d\rho}{\rho^2}
\end{equation}
with $i=1,\ldots,d-2$, corresponds to a thermal state with inverse temperature $T^{-1} = \beta = 4 \pi \rho_0 / (d-1)$. The associated expectation value for the stress tensor reads
\begin{equation}
\label{E:stresstensor}
	\frac{\hbox{Tr}\left( e^{-\beta H} T^{\mu\nu} \right)}{\hbox{Tr}\left(e^{-\beta H}\right)} = \frac{(d-1)L^{d-2}}{16 \pi G_N} \left( \frac{4\pi T}{(d-1)}\right)^{d-1}  \begin{pmatrix} d-2 &  &  & 0 \\   & 1 &  &  \\   &  & \ddots \\ 0  &  &  & 1 \end{pmatrix} \,,
\end{equation}
and the associated metric is
\begin{equation}
	g_{(0)\mu\nu} = \eta_{\mu\nu}\,.
\end{equation}
In \cite{Bhattacharyya:2007vjd} it was shown that by slightly perturbing \eqref{E:BH} in a derivative expansion one obtains a stress tensor of the form \eqref{E:constitutive} where
\begin{equation}
	P(T) = p_0 T^{d-1}\,, 
	\qquad 
	\eta = \frac{\frac{\partial P}{\partial T}}{4\pi}\,,
	\qquad 
	\zeta = 0 \,,
\end{equation}
with
\begin{equation}
	p_0 =  \frac{(d-1)L^{d-2}}{16 \pi G_N} \left( \frac{4\pi }{(d-1)}\right)^{d-1}
\end{equation}
corresponding to a conformal fluid with the universal value of the shear viscosity to entropy density ratio \cite{Kovtun:2004de}.

As discussed in the previous section, we would like to consider a thermal state of the conformal field theory which is excited by a random metric perturbation at time $t=0$, 
\begin{equation}
\label{E:g0val}
	g_{(0)\mu\nu} = - 1 \begin{cases} + 0 & t<0 \\
		- 2 \Phi \delta^t_{\mu} \delta^{t}{}_{\nu} & t> 0 
		\end{cases}\,.
\end{equation}
where $\Phi$ is a random variable given by \eqref{E:Qtoq}. 
%
%That is, from a non relativistic perspective, the thermal state is excited by a stochastic gravitational potential, $\Phi$. From the perspective of the gravitational theory, this corresponds to a solution to the Einstein equations following from \eqref{E:action} of the form \eqref{E:BH} up to $t=0$, and after which the boundary value of the metric $g_{(0)\mu\nu}$ is randomly perturbed in the time-time direction. 
After solving the equations of motion with these initial and boundary conditions, we can read off the stress tensor of the dual field theory using \eqref{E:Tmap} and infer the dynamics of the associated fluid by computing the velocity field and temperature via
\begin{equation}
	T^{\mu\nu}u_{\nu} = -(d-2)p_0 T^{d-1} u^{\mu}\,.
\end{equation}
In the remainder of this section, we will go over this procedure in detail.
We choose a coordinate system
\begin{equation}
\label{E:EFcoordinates}
\ell^{-2} ds^2 =  \Sigma(t,\vec{x},\rho)^2 \tilde{g}_{ij}(t,\vec{x},\rho) dx^{i}dx^{j}   
 	- 2 dt \left(F_{i}(t,\vec{x},\rho) dx^{i} +A(t,\vec{x},\rho)dt + \omega_0(t,\vec{x}) \frac{d\rho}{\rho^2}\right) 
\end{equation}
with $|\tilde{g}|=1$ and $i=1,\ldots,d-2$ to parameterize our metric. Any metric can be brought to the form \eqref{E:EFcoordinates} by identifying $\rho$ as a non-affine coordinate along null geodesics parameterized by the coordinates $(t,\vec{x})$.\footnote{More precisely, we consider a congruence of null geodesics all of which pass through a hypersurface $\Sigma$. We use coordinates $(t,x^i)$ on $\Sigma$ and use $\tilde{\rho}$ as an affine coordinate along the geodesics. The coordinate system \eqref{E:EFcoordinates} is obtained from the latter by stretching the radial coordinate $d\tilde{\rho} = d\rho/\rho$ and identifying an appropriate time coordinate.} Recall that the spatial coordinates $x^i$ are periodic with length $L$. %Among other things, this implies that momenta are discrete.

The stress tensor associated with \eqref{E:EFcoordinates} can be obtained by going from \eqref{E:EFcoordinates} to \eqref{E:FGsol}. We find  for $d=4$
\begin{align}
\begin{split}
\label{E:holographicTmn}
	\frac{16 \pi G_N}{3 L^2} \hbox{Tr}\left(\varrho T^{\mu\nu}\right) = &  T_h^{\mu\nu} +F^{\mu\nu}[\Phi] \ ,
\end{split}
\end{align}
where 
\begin{align}
\begin{split}
\label{E:holographicTmn}
	T_h^{\mu\nu} = &  \begin{pmatrix}
		-\frac{4 A^{(3)}}{3(1+2 \Phi)^2} & \frac{F_1^{(3)}}{1+2 \Phi}  & \frac{F_2^{(3)}}{1+2 \Phi} \\
	\frac{F_1^{(3)}}{1+2 \Phi} & \tilde g_{22}^{(3)}  -\frac{2 A^{(3)}}{3(1+2 \Phi)} &   \tilde g_{12}^{(3)}  \\
		\frac{F_2^{(3)}}{1+2 \Phi}  &\tilde g_{12}^{(3)}   &   -g_{22}^{(3)} - \frac{2 A^{(3)}}{3(1+2 \Phi)}
		\end{pmatrix}\ ,
\end{split}
\end{align}
and the superscript $(3)$ denotes an appropriate term in a series expansion near the boundary (located at $\rho = 0$), 
\begin{align}
\begin{split}
	F_i &= \frac{\partial_i \omega_0}{\rho} \iffalse- \omega_0 \partial_i \lambda \fi + F_i^{(3)} \rho + \ldots \\
	A & = \frac{1+2 \Phi}{2} \iffalse \left(\frac{1}{\rho}+ \lambda\right)^2 \fi \frac{1}{\rho^2}\iffalse- \omega_0 \partial_t \lambda \fi - \frac{1}{2} \sum_i \partial_i (\omega_0 \partial_i \omega_0) + A^{(3)}\rho + \ldots \\
	\tilde{g}_{ij} &= \delta_{ij} + \tilde{g}_{ij}^{(3)} \rho^3\,.
\end{split}
\end{align}
The explicit dependence of $F^{\mu\nu}$ on $\Phi$ is somewhat long and we omit its explicit form. Instead, we note that, with $Q=\sqrt{1+2\Phi}$
\begin{equation}
\label{E:dTdecomposition}
     \nabla_{\mu}T^{\mu}{}_{\nu} = \nabla_{\mu} T_h^{\mu}{}_\nu -  \delta^{ 0}_\nu \frac{1}{3 Q} \left((\partial_y^2 Q)^2 + 4 (\partial_x \partial_y Q)^2 - 2 \partial_x^2 Q \partial_y^2 Q + (\partial_x^2 Q)^2 -Q(\partial_y^4 Q+ 2 \partial_x^2 \partial_y^2 Q+\partial_x^4 Q)  \right)\,.
\end{equation}

We would like to solve the Einstein equations such that the asymptotic AdS boundary metric takes the form \eqref{E:g0val} and that for $t<0$ the boundary theory is in a therm state or, equivalently, the bulk geometry is a black hole. Thus, at early times, $t \leq 0$, the boundary metric is given by the black brane solution \eqref{E:BH},
\begin{equation}
\label{E:BBsolution}
	\Sigma=\frac{1}{\rho}\,,
	\qquad
	F_i = 0 \,,
	\qquad
	\tilde{g}_{ij} = \delta_{ij} \,,
	\qquad
	2 {A} = \frac{1}{\rho^2}  -\frac{\rho^{1}}{\rho_0^{3}}\,,
	\qquad
	\omega_0 = 1\,.
\end{equation}
The associated stress tensor is given by \eqref{E:stresstensor}.
For later times we need to solve the Einstein equations numerically. We do so following an algorithm similar to that of \cite{Chesler:2013lia}. We replace all time derivatives with derivatives along the null direction\,,
\begin{equation}
	d_+ = \partial_t - \frac{\rho^2}{\omega_0}A \partial_{\rho} \ .
\end{equation}
With these coordinates the equations of motion take the form of a set of nested linear equations. The expression for $\omega_0$ is given by
\begin{equation}
	\omega_0^2 = 1+2 \Phi\,.
\end{equation}

Since some of the metric components diverge near the asymptotic boundary, it is convenient to work with variables which are sufficiently regular there:
\begin{align}
\begin{split}
\label{E:subtraction}
	\Sigma=& \frac{1}{\rho}\iffalse+\lambda\fi +\rho^3 \widehat{\Sigma} \\
	F_i =& \frac{\partial_i  \omega_0}{\rho}\iffalse- \omega_0\, \partial_i \lambda\fi+ \widehat{F_i}\\
	A = & \frac{1+2 \Phi}{2} \iffalse\Big(\frac{1}{\rho}+\lambda \Big)^2\fi \frac{1}{\rho^2} \iffalse-\omega_0\, \partial_t \lambda\fi-\frac{1}{2}\Big(\sum_i \partial_i(\omega_0 \partial_i \omega_0)\Big)
	 +  \widehat{A}\\
	 d_+\sigma = &  \frac{1+2\Phi}{2} \iffalse \Big(\frac{1}{\rho}+\lambda \Big)^2\fi \frac{1}{\rho^2}-\frac{1}{2 \omega_0}\Big(\sum_i \partial_i(\omega_0 \partial_i \omega_0)\Big)+\rho \,\widehat{d_+\sigma} \\
	d_+ \tilde{g}_{ij}=&\rho^2\widehat{d_+ \tilde{g}_{ij}} ,
\end{split}
\end{align}
where the $\mathcal{O}(\rho^0)$ behavior of $\Sigma$, $F_i$ and $A$ has been chosen by fixing a residual radial coordinate reparameterization invariance. The boundary conditions for the hatted variables are given by
\begin{align}
\begin{split}
\label{E:subtraction}
	\widehat{\Sigma}|_{\rho=0}=& 0,\, \partial_\rho \widehat{\Sigma}|_{\rho=0}= 0 \\
	\widehat{F_i}|_{\rho=0}=& 0,\, \partial_\rho \widehat{F_i}|_{\rho=0}= F_i^{(3)} \\
	\widehat{A}|_{\rho=0}=& 0,\, \partial_\rho \widehat{A}|_{\rho=0}= A^{(3)} \\
	 \widehat{d_+\sigma} |_{\rho=0} = & A^{(3)}\\
	\widehat{d_+ \tilde{g}_{ij}}|_{\rho=0} =& -\frac{3}{2}\tilde{g}_{ij}^{(3)}.
\end{split}
\end{align}
where $F_i^{(3)}$, $A^{(3)}$ and $\tilde{g}_{ij}^{(3)}$ need to be supplied initially. The expressions for  $F_i^{(3)}$ and $A^{(3)}$ can be obtained at later times from the evolution equation for the stress tensor
\begin{equation}
	\label{E:covariantDivergence}
 \nabla_{\mu}T^{\mu\nu}=0
\end{equation}
where $T^{\mu\nu}$ is given by \eqref{E:holographicTmn} and we make use of \eqref{E:dTdecomposition} while $\tilde{g}_{ij}^{(3)}$ can be obtained by integrating $\widehat{d_+ g_{ij}}$,
\begin{equation}
\partial_t \tilde{g}_{ij} = d_+ \tilde{g}_{ij}+\frac{\rho^2}{\omega_0} A \, \partial_\rho \tilde{g}_{ij}\,.
\end{equation}

\section{Numerical Results}
\label{E:Results}
 We solve the Einstein equations discretized on a  spatial Fourier grid with grid size $N_x= N_y=70$ and a radial Chebyshev grid with $N_\rho=11$ grid points. As a consequence of using periodic (Fourier) cardinal functions to approximate the spatial dependence of all functions, our spatial integration domain is a 2-torus, whose physical size we choose to be $L_x = L_y = 250 \rho_0 $. The parameter $\rho_0$ is associated with the initial conditions for the geometry which are specified by the black hole geometry \eqref{E:BH}. Recall that $\rho_0$ is related to the equilibrium  temperature through $\rho_0 = \frac{3}{4\pi T_0}$ 
 on the initial time slice. We set $D= 1  \times 10^{-6} \rho_0$, $\tau = 0.3 \, \rho_0$ choose a time stepping size of $\delta t = 0.025 \,\rho_0$ a forcing scale of $k_f =  \frac{42 \pi}{L}$ and an  annulus size of  $\Delta k_f =  \frac{6\pi}{4L}$.  During the time evolution we employ a 2/3-filter, removing the largest third of modes with a smooth cut-off.  We repeated the evolution of the geometry under the influence of a randomly flactuating $tt$-component of the boundary metric $24$ times, to create an ensemble in order to compute ensemble averages. 
We present the results for the calculation of various fluid observables at the boundary in the following subsections.

\subsection{Fluid velocity and vorticity}
To compare our numerical data at the boundary with the hydrodynamic data we can identify the velocity field and energy density (in the Landau frame) from the stress tensor by solving the eigenvalue equation
 \begin{equation}
	T^\mu_\nu u^\nu = -e u^{\mu},
 \end{equation}
The components of the fluid velocity $v_1$ and $v_2$ are simply
\begin{equation}
	u^{\mu} = \frac{1}{\sqrt{1-v_1^2-v_2^2}} \left(1,v^1,v^2\right)^{\mu}\,.
\end{equation}
We will denote the nonrelativistic velocity vector $\vec{v} = (v^1,v^2)$. Our runs are such that $|\vec{v}| \lesssim 8 \times 10^{-3} c$ so we are well within the nonrelativistic limit. (Here $c$ is the speed of light, inserted in this instance for clarity.)

\begin{figure}[h]
	\includegraphics[scale=0.45]{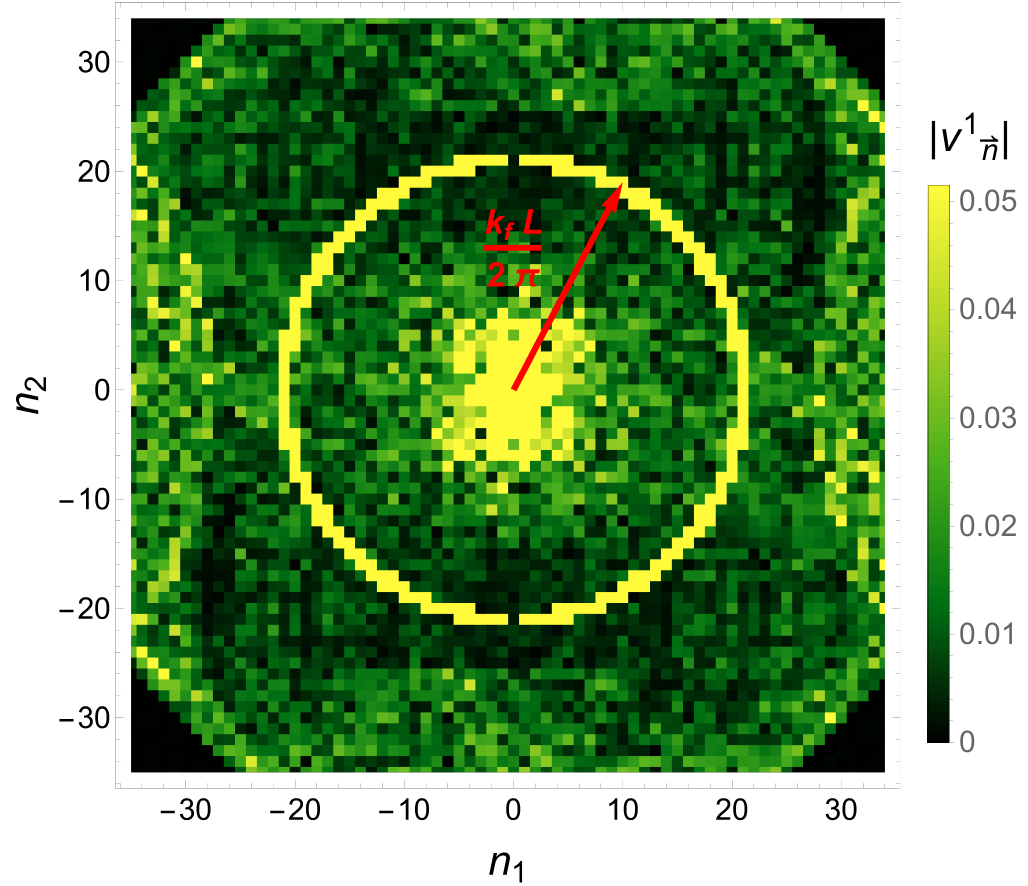}
	\hfill
	\includegraphics[scale=0.45]{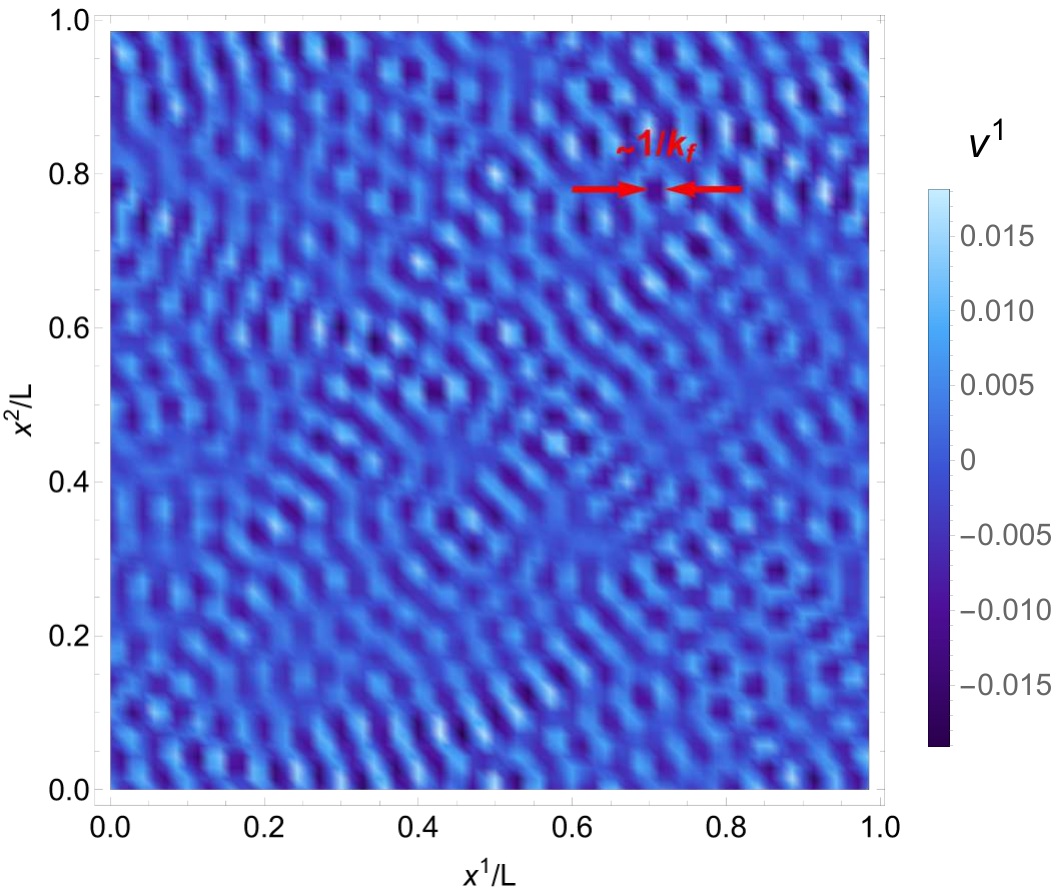}
	\caption{Plots of the nonrelativistic velocity field $v^1$ in Fourier space (left) and real space (right) demonstrating the dominant contribution of the driving force at momentum scales of order $k_f$.}
	\label{F:Fourierv1}
\end{figure}

On the left panel of figure \ref{F:Fourierv1} we display a snapshot of the Fourier components of the fluid velocity (for a single run),
\begin{equation}
	\vec{v}(\vec{x}) =\sum_{\vec{n}} \vec{v}_{\vec{n}} e^{i \frac{2\pi \vec{n}}{L} \cdot \vec{x}}
 \end{equation}
at late times, $t=550 \rho_0$. 
Clearly, the velocity field is mainly supported at wavelengths around $k_f$. A real space plot of the velocity field, given in the right panel of figure \ref{F:Fourierv1} displays the same information.

\begin{figure}[h]
	\includegraphics{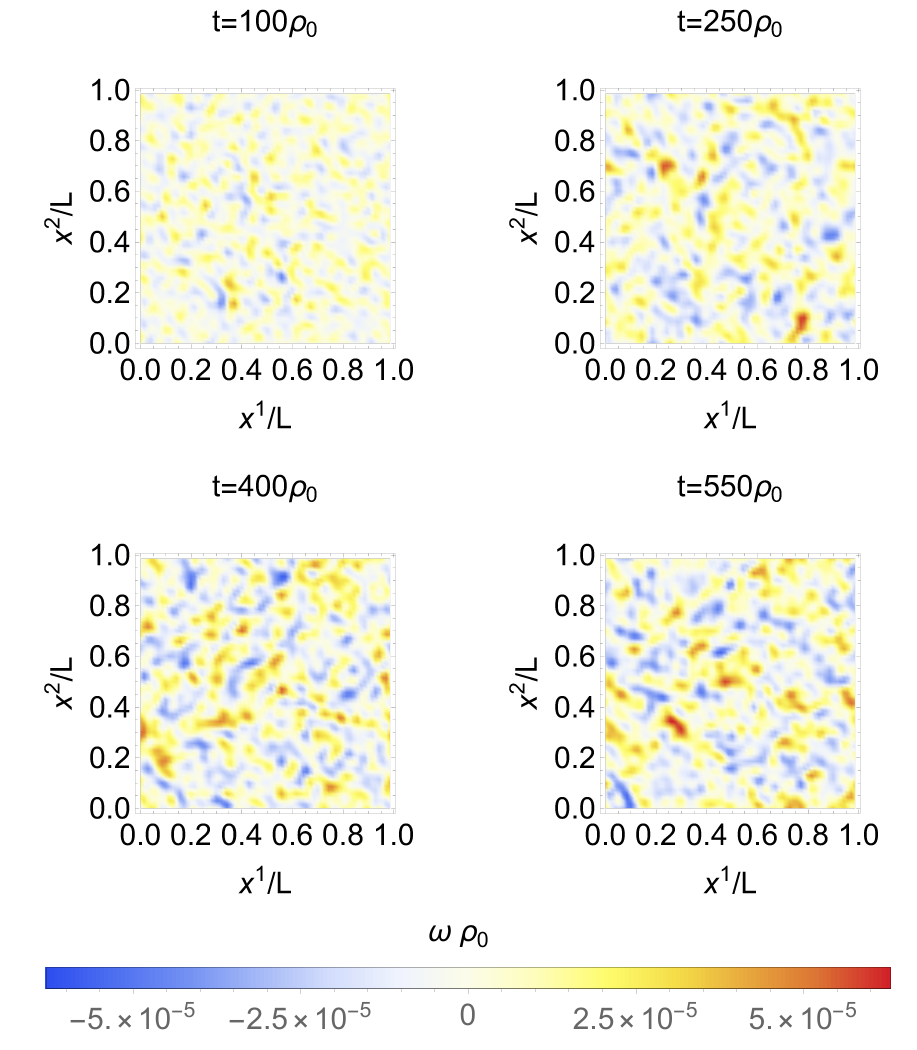}
	\caption{The vorticity for the filtered velocity as a function of time. As time evolves larger vortices are created as expected.}
	\label{F:vorticity}
\end{figure}

In order to see the fluid structure underlying that determined by the driving force we filter out all wavelengths above $k_f$. %The resulting velocity field is depicted in figure \ref{F:v1wokf}. 
An improved visual depiction of the fluid can be found in figure \ref{F:vorticity}  where we plot the (filtered) vorticity as a function of time. 

 \subsection{Compressiblity}
Recall that in section \ref{S:NRlimit} we argued that the nonrelativistic potential flow resulting from a random gravitational potential will be compressible. To see that this is indeed the case we define $C = \overline{\left(\nabla_i v^i\right)^2} / \overline{\left(\epsilon^{ij} \nabla_i v_j\right)^2} $ as a measure for compressibility. In the strictly incompressible limit we expect that $C=0$ by the equations of motion. In the incompressible limit we expect that $C  = \mathcal{O}(\epsilon^2)$ where $\epsilon$ is the magntiude of the velocity field or, if the flow is incompressible, the magnitude the root of the fluctuations of the potential. In the runs described above we found that for $t>100$,
\begin{equation}
	C \sim 5400,\,
	\quad
	\overline{|\vec{v}|^2} \sim 7 \times 10^{-5},\,
	\quad
	\overline{\phi^2} \sim 4.167 \times 10^{-7}\,,
\end{equation}
indicative of the compressible nature of the flow.

\begin{figure}[h]
\begin{center}
	\includegraphics[scale=0.45]{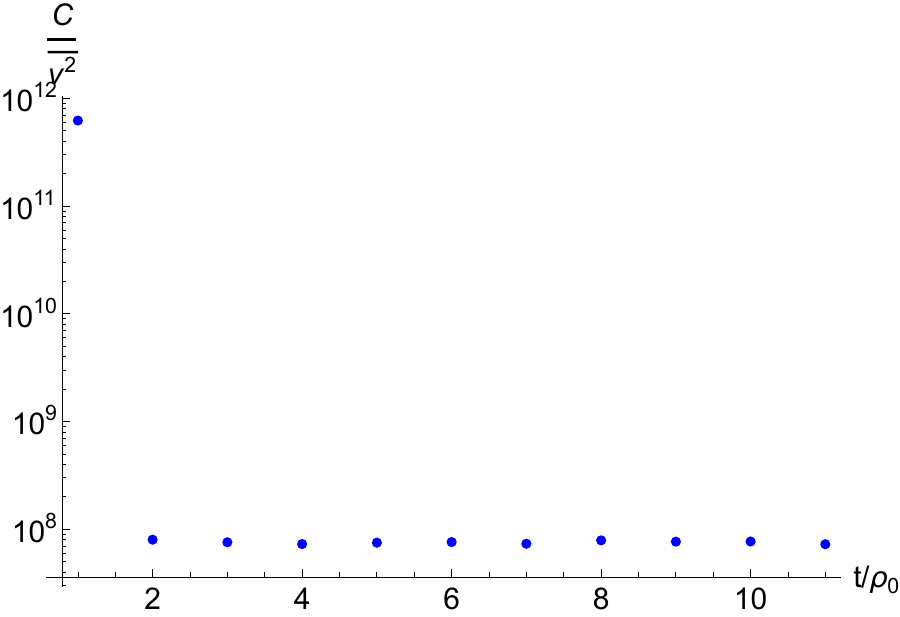}
	\hfill
	\includegraphics[scale=0.45]{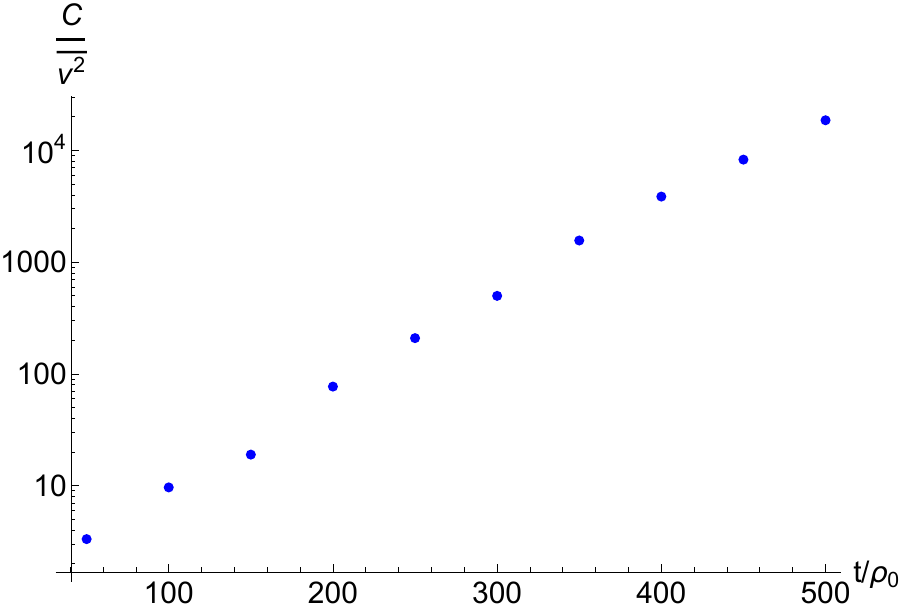}
	\caption{A plot of the measure of compressability, $C =\overline{\left(\nabla_i v^i\right)^2} / \overline{\left(\epsilon^{ij} \nabla_i v_j\right)^2} $ as a function of time compared to the average velocity squared for different initial conditions. On the left the initial conditions where that of a stationary thermal state, inline with the other plots shown in this section. On the right we consider an additional run where the initial conditions where that of an incompressible fluid. We have stopped the run at $t=500 \rho_0$. For strictly incompressible flow the value of $C$ should be zero and in the incompressible limit of relativistic flow the value of $C$ should be of order $|\vec{v}|^2$ (in units of the speed of light).}
	\label{F:compressibility}
\end{center}
\end{figure}

As mentioned earlier, the system we have in mind is in a thermal state at $t=0$ for which $C$ is not well defined since the velocity field vanishes. As the forcing is turned on, $C$ approaches $5400$ after, roughly $t=100 \rho_0$. One might worry that compressibility or the lack of it may be a result of a poor choice of initial conditions. To argue that this is not the case we studied systems whose initial velocity configuration $\vec{v} \neq 0$ satisfied the incompressibility condition $\vec{\nabla}\cdot \vec{v} = 0$ and let the system evolve in time. In figure \ref{F:compressibility} we see that the compressibility parameter increases exponentially indicating that the system prefers to reach compressible flow, inline with the discussion in section \ref{SS:NRL}.

 \subsection{Measures of turbulence}
While the flow is compressible we can still compare the properties of the solutions we obtain to those of incompressible turbulent flow. The hallmark of incompressible turbulent flow is the Kolmogorov power law for the kinetic energy power spectrum per unit mass, 
 \begin{equation}
 \label{E:PS}
	P_v(k) = \frac{1}{2} \int  | \hat{\vec{v}} (k') |^2 \delta(|k'|-|k|)d^2k'.
 \end{equation}

\begin{figure}[h]
\begin{center}
%	\includegraphics[scale=0.6]{powerspectrum_boundary.pdf}
%	\includegraphics[scale=0.55]{powerspectrum_horizon.pdf}
%\caption{On the left: the double logarithmic plot of the ensemble averaged energy power spectrum $E(k)$ computed at the boundary as a function of the wave vector $k$ at various time steps, compared with the expected $k^{-5/3}$ scaling (red dotted line). On the right: the analogous plot for the energy power spectrum  computed from the fluid velocity on the horizon as defined in appendix A.  At time $t=500 \, \rho^{-1}_h$ the linear fit to the log-log plot of the power spectra yields a scaling of $k^{-1.82}$ for the power spectrum computed on the boundary and a scaling of $k^{-1.91}$ for the power spectrum computed on the horizon.}
	\includegraphics[scale=0.44]{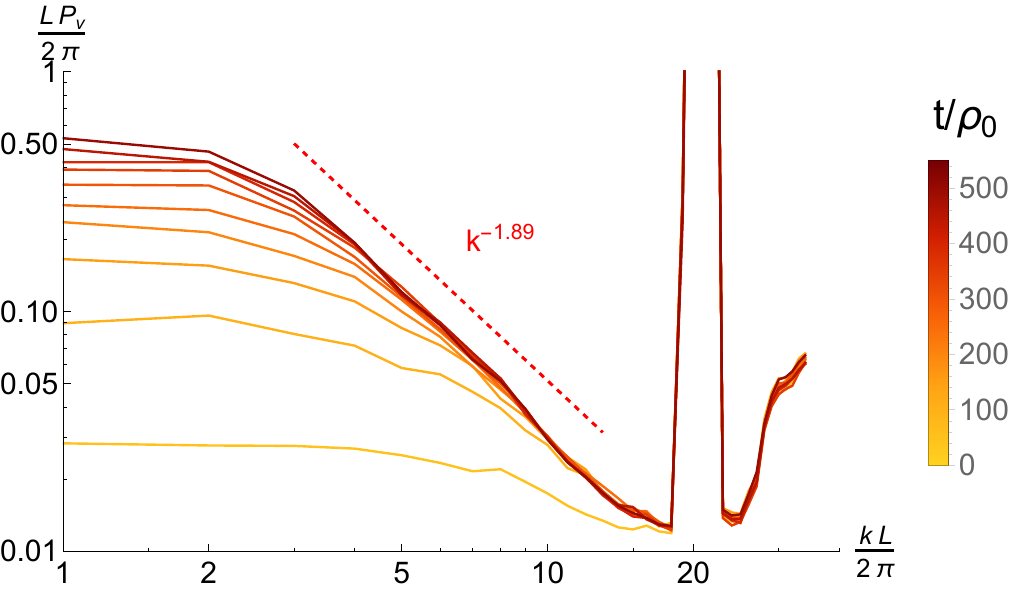}
	\hfill
	\includegraphics[scale=0.44]{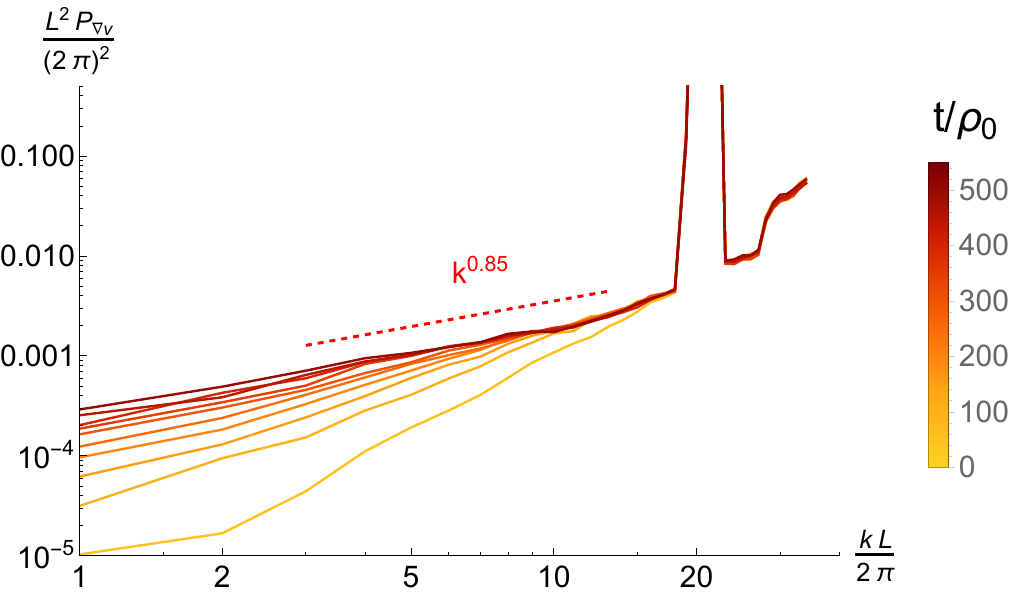}
	\hfill
	\includegraphics[scale=0.44]{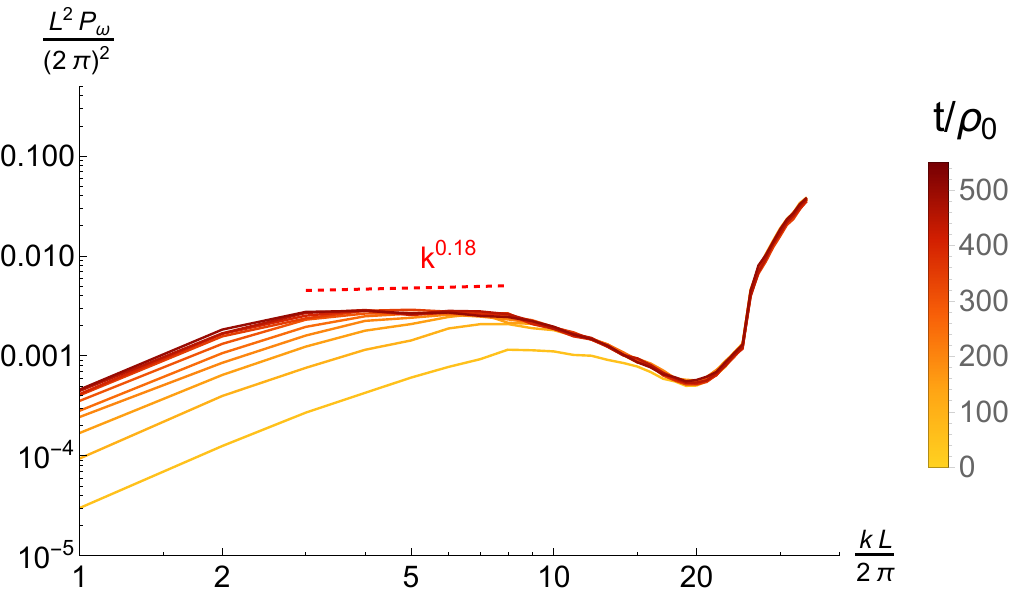}
	\caption{Power spectra for the velocity field (top left), expansion (top right), and vorticity (bottom) of the fluid (see \eqref{E:PS} and \eqref{E:otherPS}) as a function of time. The power spectrum for very early times is below the threshold of the plot. The peak at $k_f = 2 \pi/L \times 21$ in the top plots is a result of the irrotational driving force. The increase in the power spectrum at $k>k_f$ is likely a result of the relativistic origin of the fluid and related to the appearance of higher harmonics of the driving force. A fit of the power spectrum to a power law $k^s$ over a region specified by the dashed red line gives $s = -1.90 \pm 0.06$, $s=0.8\pm 0.1$ and $s=0.1 \pm 0.3$ for the velocity, expansion and vorticity power spectra respectively. The dip in the power spectrum of the vorticity at momenta below $k_f$ is similar to that appearing in \cite{Kritsuk_2019}.} 
\label{F:power_spectrum_fig}
\end{center}
\end{figure}

In figure \ref{F:power_spectrum_fig} we plot of the ensemble averaged power spectrum as a function of the wave vector $k$ at various time steps. %It seems to agree with a $k^{-2}$ scaling which appears in the , compared with Kolomogorov's $k^{-5/3}$ scaling (red dotted line).
%The left and right plots are the calculations on the boundary and the horizon respectively.
While the flow is not, strictly speaking, incompressible, the time evolution of the energy power spectrum does seem to lead to features similar to the K41 theory. A fit of the power law data to a power law behavior in the interval $3 \leq \frac{L k}{2\pi} \leq 14$ gives an exponent, $k^{s}$, of $s = -1.9 \pm 0.06$ in agreement with similar results for two (spatial) dimensional driven compressible flow \cite{PhysRevE.75.046301,Kritsuk_2019} where $P_v \sim k^{-2}$. %The difference between $s=-2$ and $s=-5/3$ of the Kraichnan theory is sometimes attributed to the absence of large scale dissipation. 

In order to better understand the contribution of the solenoidal and irrotational component of the velocity fields to $P_v$ it is convenient to decompose the velocity field into a solenoidal (incompressible) component and an irrotational (compressible) component, $\vec{v} = \vec{v}_i + \vec{v}_c$ as in \eqref{E:udecomposition}. The expansion and vorticity of the fluid velocity are associated with the compressible and incompressible components of $\vec{v}$ respectively,
\begin{align}
\begin{split}
	\vec{\nabla} \cdot \vec{v} = \vec{\nabla} \cdot \vec{v}_c\,,
	\qquad
	\omega = \vec{\nabla} \times \vec{v} = \vec{\nabla}\times \vec{v}_i\,.
\end{split}
\end{align}
Here $\vec{\nabla} \times \vec{v} = \epsilon^{ij} \partial_i v_j$. In addition to the power spectrum for the velocity field we have computed the power spectrum of the expansion and vorticity,
\begin{align}
\begin{split}
\label{E:otherPS}
	P_{\nabla v} &= \int | \widehat{\nabla \cdot v}(k') |^2 \delta(|k'|-|k|)d^2k'\,, \\
	P_\omega(k) &= \int  | \hat{\omega} (k') |^2 \delta(|k'|-|k|)d^2k',
\end{split}
\end{align}
from which we can infer the power spectrum for the compressible and incompressible components of the velocity field.
%\YO{Let's discuss the relationship with the icompressible flow with two cascades}\AY{I made some modifications %to the text below}
We have found that $P_{\nabla v} \sim k^{s}$ with $s \sim 0.8 \pm 0.1$ in the inertial range implying that $P_{v_c} \sim  k^{-1.2}$ while $P_{\omega} \sim k^s$ with $s = 0.1 \pm 0.3$ implying that $P_{v_i} \sim k^{-1.9}$ so that the incompressible component of the velocity field dominates the power spectrum (see figure \ref{F:power_spectrum_fig}). The latter is compatible with  with \cite{Kritsuk_2019} who studied compressible flow with a solenoidal driving force also in the absence of large scale friction (see also \cite{PhysRevE.75.046301}). The power spectrum for the (sub-dominant) compressible component of the velocity field we obtained differs from that of \cite{Kritsuk_2019} who obtained $P_{\omega} \sim k^{0.2}$ possibly due to the nature of the driving force used in these simulations (solenoidal in \cite{Kritsuk_2019} vrs. rotational driving in our setup).  

To summarize our findings so far, while our flow is compressible, as indicated in figures \ref{F:compressibility}, the power spectrum for the velocity field is dominated by its incompressible component. The deviation of the (incompressible) velocity field power spectrum from the $-5/3$ power law behavior may be attributed to the lack of large scale friction \cite{PhysRevE.75.046301} which results in coherent vortices.
 
Another measure of turbulence are the moments of the local energy dissipation. For incompressible flow, the local energy dissipation is given by
\begin{equation}
\label{epsilon}
	\epsilon(x) =  \frac{\nu}{2}  \left(\partial_i v^j +  \partial_j v^i \right)^2 \,.
\end{equation}
This is the rate at which kinetic energy of the fluid is lost to friction. The space averaged local energy dissipation over a ball of radius $r$, $B(r)$, centered around $x$ reads
\begin{equation}
\label{mea}
	\epsilon_r(x) = \frac{1}{Vol(B_d(r))} \int_{|x'-x| \leq r} d^d x'  \epsilon(x') \,.
\end{equation}
Under Kolmogorov's theory (K41) \cite{Kolmogorov1941}, the local energy dissipation is expected to be uniform in the inertial range so that $\overline{\epsilon_r(x)}$ should be independent of $r$ (and $x$) for radii in the inertial range.

\begin{figure}
\begin{center}
	\includegraphics[scale=0.44]{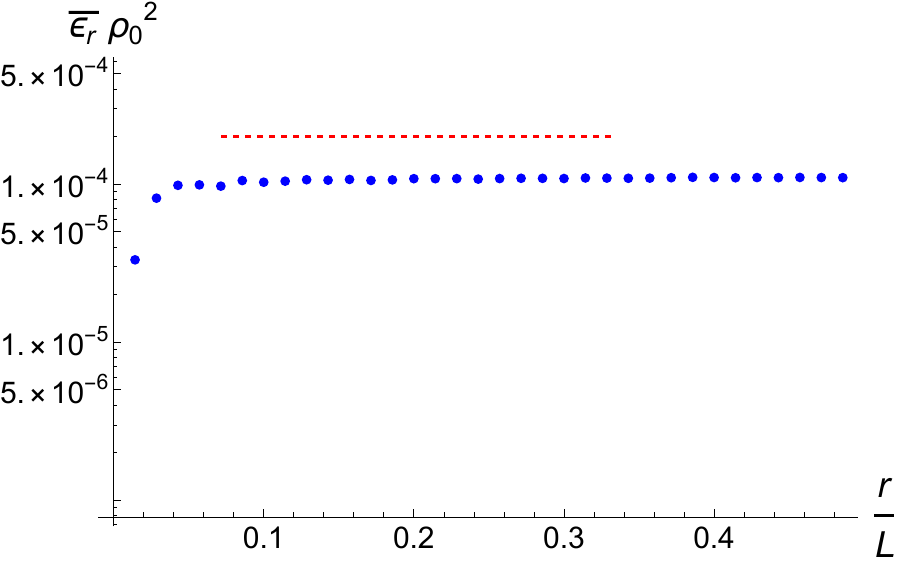}
	\hfill
	\includegraphics[scale=0.44]{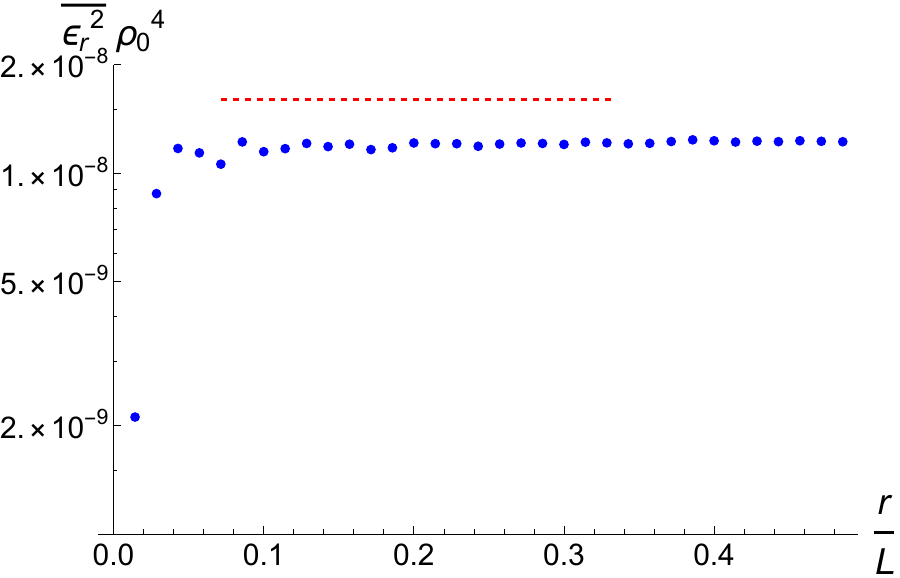}
	\caption{First (left) and second (right) moments of the averaged local energy dissipation $\epsilon_r$ as a function of $r$ at $t=550 \rho_0$ are depcited in blue. The dashed horizontal red line specifies the inertial range compatible with the region where a Kolmogorov power spectrum was computed. According to Kolmogorov's theory, the averaged energy dissipation rate should be constant in the inertial range.}
%	
%	. As expected for two dimensional turbulence,  $\epsilon_r$ shows no scaling in the inertial range, indicated by the red dotted line. On the right we display the correlation $\langle \epsilon_r \epsilon_R\rangle$ as a function of the ratio $r/R$ computed on the boundary (blue solid line). Again the red dotted line represnts the range of scales in which we observe an approximate $-5/3$-scaling for the energy power spectrum.} 
\label{F:energy_dissipation}
\end{center}
\end{figure}

In figure \ref{F:energy_dissipation} we plotted the ensemble averaged energy dissipation over a range of spheres of radius $r$. It seems that the (incompressible) averaged local energy dissipation is almost uniform in all cases inline with the Kolmogorov's theory.

\begin{figure}
\begin{center}
	\includegraphics[scale=1]{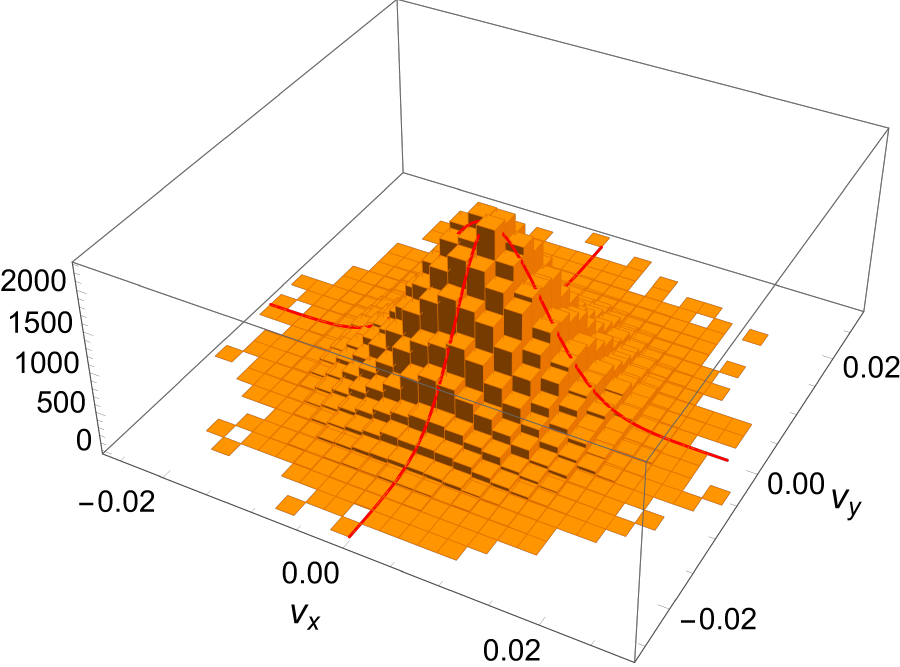}
\caption{A late time histogram of the velocity field with 28 bins in each direction. The red curves correspond to a Gaussian fit as expected in turbulence.} 
\end{center}
\label{F:Gaussian}
\end{figure}

%In \cite{Benzi_1998} it was suggested that the averaged correlator of the energy dissipation rate scale with the ratio of scales at which the energy dissipation is measured. In the current language this amounts to the dependence of $\overline{\epsilon_r \epsilon_R}$ on the ratio $r/R$. In figure \ref{F:rRplots} we plot $\overline{\epsilon_r \epsilon_R}$ as a function of $r$ and $R$. Our current data does not allow for a conclusive statement regarding this. \AY{Maybe we should not discuss it then?}
%\begin{figure}
%\begin{center}
%	\includegraphics[scale=1]{rRplot.pdf}
%\caption{A plot of the correlator of the integrated energy power spectrum $\epsilon_r$ defined in \eqref{mea}. The average has been carried out over 24 ensembles and 7 time intervals. The inertial range where a power law for the energy power spectrum was observed is marked by a dashed red line. The right plot is a zoomed in version of the left. The resolution of our data in the inertial range (marked with a dashed red line) makes the result inconclusive.} 
%\end{center}
% \label{F:rRplots}
% \end{figure}
A further indication that the flow we find is turbulent can be obtained by studying the statistical distribution of the velocity field. Within the resolution of our numerics, a Gaussian distribution provides an excellent fit to the velocity distribution, as exhibited in figure \ref{F:Gaussian}

\section{Summary and Discussion}
\label{E:Discussion}
Holography offers a novel framework to study fluid dynamics formulated via a random black hole bulk geometry.
The complex structure of fluid turbulence is encoded both on the black hole horizon as well as at the gravitational background boundary. Bulk random geometry can be generated by a random matter field or by a random gravitational metric.
In this work we used the latter and included a random gravitational potential. We showed that the small velocity limit
of our holographic model captures the dynamics of a nonrelativistic compressible fluid in the two-dimensional 
inverse cascade regime. We calculated
several fluid observables at high Reynolds number that exhibit inertial range scaling characteristics similar to those expected from Kolmogorov's theory 
of incompressible fluid turbulence.
In a future work we plan to modify the random gravitational metric, such that the incompressible fluid limit can be achieved and Kolmogorov's theory can be made holographically precise.

\section*{Acknowledgements}

The authors are grateful to L. Lehner, R. Pandit, P. Kovtun, M. Disconzi, A. Ori and A. Frishman for useful discussions. %YO, SW and AY are supported in part by an Israeli Science Foundation excellence center grant 2289/18.
YO and AY are supported in part by a binational science foundation grant 2022110, YO is supported in part by Israel Ministry of Science, SW and AY are supported in part by an ISF-NSFC grant 3191/23.

\bibliographystyle{JHEP}
\bibliography{turbulence}

\end{document}